\documentclass[10pt]{article}
\usepackage{amsmath}
\usepackage{latexsym}
\usepackage{amssymb}
\usepackage{amsfonts}
\usepackage{epsfig}
\usepackage{times}

\newtheorem{theorem}{Theorem}

\newtheorem{lemma}{Lemma}

\def\qed{\hfill{$\Diamond$} \\}

\def\expect{{\mathbb  E}}

\def\Pr{{\mathbb P}}
\def\real{{\mathbb  R}}

\def\proof{\noindent{\bf Proof:} }
\def\eqdef{\triangleq}

\begin{document}


\date{\small October 2007; revised August 2008; finalized March 2009}

\author{
{\sc Predrag R. Jelenkovi\'c} 
\thanks{Predrag Jelenkovi\'c, Department of Electrical Engineering, Columbia University, New York, NY 10027, predrag@ee.columbia.edu.}
\\ 
{\small Department of Electrical Engineering}  \\
{\small Columbia University, New York}
\and
{\sc Ana Radovanovi\'c}
\thanks{
Corresponding author: Ana Radovanovi\'c, Google Inc., New York, NY 10011, anaradovanovic@google.com.
} \\ 
{\small Google, Inc.}\\
{\small New York}
}

\title{\vspace{-0.3in}\Large Asymptotic Optimality of the Static Frequency
Caching in the Presence of Correlated Requests}

\maketitle

\normalsize

\thispagestyle{empty}

\begin{abstract}
It is well known that the static caching algorithm that keeps the most
frequently requested documents in the cache is optimal in case when documents
are of the same size and requests are independent and equally distributed.
However, it is hard to develop explicit and provably optimal caching algorithms when
requests are statistically correlated. In
this paper, we show that keeping the most frequently requested
documents in the cache is still optimal for large cache sizes even
if the requests are strongly correlated. 

\vspace{12pt} \noindent{\bf Keywords:} Web caching, cache fault
probability, average-case analysis, least-frequently-used caching,
least-recently-used caching, long-range dependence
\end{abstract}

\section{Introduction}
\label{sec:intro}
One of the important problems facing current and future network
designs is the ability to store and efficiently deliver a huge
amount of multimedia information in a timely manner. Web caching
is widely recognized as an effective solution that improves the
efficiency and scalability of multimedia content delivery,
benefits of which have been repeatedly verified in practice. For
an introduction to the concept of Web caching, the most recent
tutorials, references and the latest technology, an interested
reader is referred to the Web caching and content delivery Web
page \cite{BDxx}.

Caching is essentially a process of storing information closer to
users so that Internet service providers, delivering a given
content, do not have to go back to the origin servers every time
the content is requested. It is clear that keeping more popular
documents closer to the users can significantly reduce the traffic
between the cache and the main servers and, therefore, improve the
network performance, i.e., reduce the download latency and network
congestion. One of the key components of engineering efficient Web
caching systems is designing document placement/replacement
algorithms (policies) that are managing cache content, i.e.,
selecting and possibly dynamically updating a collection of cached
documents.

 The main performance objective in creating and implementing these algorithms is
minimizing
 the long-term fault probability, i.e., the average number of misses
during a long time period. In the context of equal size documents
and independent reference model, i.e., independent and identically
distributed requests, it is well known (see \cite{OBAM03},
Chapter~6 of \cite{COD73}) that keeping the most popular documents
in the cache optimizes the long term cache performance; throughout
this paper we refer to this algorithm as {\it static frequency
caching}. A practical implementation of this algorithm is known as
Least-Frequently-Used rule (LFU). However, the previous model does
not incorporate any of the recently observed properties of the Web
environment, such as: variability of document sizes, presence of
temporal locality in the request patterns (e.g.,
 see \cite{JERA02a}, \cite{JB00}, \cite{ABC96}, \cite{BCF99}, \cite{CAI97} and references
 therein), variability in document popularities (e.g., see
 \cite{AW96}) and retrieval latency (e.g., see \cite{WA97}).

Many heuristic algorithms that exploit the previously mentioned
properties of the Web environment have been proposed, e.g., see
\cite{CAI97}, \cite{OBAM03}, \cite{JB00a} and references therein.
However, there are no explicit algorithms that are provably
optimal when the requests are statistically correlated even if
documents are of equal size. Our main result of this paper, stated
in Theorem~\ref{theorem:main} of Section~\ref{sec:main}, shows
that, in the generality of semi-Markov modulated
requests, the static frequency caching algorithm is still optimal
for large cache sizes. The semi-Markov modulated processes,
described in Section~\ref{sec:model}, are capable of modeling a
wide range of statistical correlation, including the long-range
dependence (LRD) that was repeatedly experimentally observed in
Web access patterns; these types of models were recently used in
\cite{JERA02} and their potential confirmed on real Web traces in
\cite{JERA02a}. In Section~\ref{sec:concl}, under mild additional
assumptions, we show how our result extends to variable page
sizes. Our optimality result provides a benchmark for evaluating
other heuristic schemes, suggesting that any heuristic caching
policy that approximates well the static frequency caching should
achieve the nearly-optimal performance for large cache sizes. In
particular, in conjunction with our result from \cite{JERA02}, we
show that a widely implemented Least-Recently-Used (LRU) caching
heuristic is, for semi-Markov modulated requests and generalized
Zipf's law document frequencies, asymptotically only a factor of
1.78 away from the optimal. 
Furthermore, similar results can be expected to hold 
for the improved version of the LRU caching, termed Persistent Access Caching,
that was recently proposed and analyzed in \cite{JERA08}.

 \section{Modeling statistical dependency in the request process}
 \label{sec:model}

In this section we describe a semi-Markov modulated request
process. As stated earlier, this model is capable of capturing a
wide range of statistical correlation, including the commonly
empirically observed LRD. This approach was recently used in
\cite{JERA02}, where one can find more details and examples.

 Let a sequence of requests arrive at Poisson points $\{\tau_{n},-\infty
<n<\infty\}$ of unit rate.
 At each point $\tau_{n}$, we use $R_{n}$, $R_{n}\in \{1,2,\dots ,N\}$, to
 denote a document that  has been requested, i.e., the event
$\{R_{n}=i\}$ represents a request for document $i$ at time
 $\tau_{n}$; we assume that the sequence $\{R_{n}\}$ is independent of
the arrival Poisson points
 $\{\tau_{n}\}$ and that $\Pr[R_n=i]>0$ for all $i$ and $\Pr[R_n<\infty]=1$.

Next, we describe the dependency structure of the request sequence
$\{R_n\}$. We consider the
 class of finite-state, stationary and ergodic semi-Markov processes
$J$, with jumps at almost
 surely strictly increasing points $\{T_{n},-\infty <n<\infty \}$,
$T_{0}\le 0< T_{1}$. Let process $\{J_{T_{n}},-\infty<n<\infty\}$
be an irreducible Markov chain that is independent of $\{\tau_n\}$, has finitely many
 states $\{1,\dots,M\}$ and transition matrix $\{p_{ij}\}$. Then, we construct a piecewise constant and right-continuous {\it modulating
process} $\{J_t\}$ such that \\
 \begin{equation*}
 J_{t}=J_{T_{n}},\;\;\;\text{if $T_{n}\le t<T_{n+1}$};
 \end{equation*}
for more details on the construction of process
 $J_{t}$, $t\in {\mathbb R}$ see Subsection 4.3 of
\cite{JERA02}.
 Let $\pi_{r}=\Pr [J_{t}=r]$, $1\le r\le M$, be the stationary
distribution of $J$ and, to avoid trivialities, we assume that
$\min_{r}\pi_{r}>0$. For each
 $1\leq r\leq M$, let $q_i^{(r)},1\leq i\leq N\leq \infty$, be a
probability mass function, where $q_{i}^{(r)}$ is
 used to denote the probability of requesting item $i$ when the
underlying process $J$ is in state
 $r$. Next, the probability law of $\{R_{n}\}$ is uniquely determined by the
modulating process $J$ according
 to the following conditional distribution,
 \begin{equation}
 \label{eq:condindep}
 \Pr [R_{l}=i_{l},1\le l\le n|J_{t},0\le t\le
\tau_{n}]=\prod_{l=1}^{n}q_{i_{l}}^{(J_{\tau_{l}})},\;\;\;\;\;n\geq
1,
 \end{equation}
 i.e., the sequence of requests $R_{n}$ is conditionally independent
given the modulating
 process $J$. Given the properties introduced above, it is easy to
conclude that the constructed
 request process $\{R_{n}\}$ is stationary and ergodic as well. We will use
 \begin{equation*}
  q_{i}=\Pr [R_n=i]=\sum_{r=1}^{M}\pi_{r}q_{i}^{(r)}
 \end{equation*}
 to express the marginal request distribution, with the assumption
 that $q_{i}>0$ for all $i\ge 1$. In addition, assume that requests
 are enumerated according to the non-increasing order of marginal
 request popularities, i.e., $q_{1}\ge q_{2}\ge \dots$.

In this paper we are using the following standard notation. For
any two real functions $a(t)$ and $b(t)$ and fixed $t_{0}\in
\real\cup\{\infty\}$ we use $a(t)\thicksim b(t)$ as
$t\rightarrow t_{0}$ to denote $\lim_{t\rightarrow t_{0}}
[a(t)/b(t)]=1$. Similarly, we say that $a(t)\gtrsim b(t)$ as
$t\rightarrow t_{0}$ if $\lim\inf_{t\rightarrow t_{0}}a(t)/b(t)\ge
1$; $a(t)\lesssim b(t)$ has a complementary definition, i.e.,
$\lim \sup_{t\rightarrow t_{0}} a(t)/b(t) \le 1$.

Throughout the paper we will exploit the renewal (regenerative)
structure of the semi-Markov process.
 In this regard, let $\{{\cal T}_{i}\}$, ${\cal T}_{0}\le 0 < {\cal
T}_{1}$, be a subset of points
 $\{T_{n}\}$ for which $J_{T_{n}}=1$. Then, it is well known that
$\{{\cal T}_{i}\}$ is a renewal
 process and that sets of variables
 $\{J_t, {\cal T}_j\leq t < {\cal T}_{j+1}\}$ are independent for
different $j$ and identically distributed, i.e.,  $\{{\cal
T}_{i}\}$
 are regenerative points for $\{J_t\}$. Furthermore, the conditional
independence
 of $\{R_n\}$ given $\{J_t\}$, implies that $\{{\cal T}_{i}\}$
 are regenerative points for $R_n$ as well.

Next we define ${\cal R}_r(u,t)$, $1\le r\le M$, to be a set of
distinct requests that arrived in interval $[u,t)$, $u\le t$, and
denote by $N_{r}(u,t)$, $1\le r\le M$, the number of requests in
interval $[u,t)$ when process $J_{t}$ is in state $r$.
Furthermore, let $N(u,t)\eqdef N_{1}(u,t)+\dots +N_{M}(u,t)$
represent the total number of requests in $[u,t)$; note that
$N(u,t)$ has Poisson distribution with mean $t-u$.

The following technical lemma will be used in the proof of the
main result of this paper.

\begin{lemma}
\label{lemma:asympr} For the request process introduced above, the
following asymptotic relation holds
\begin{equation}
\label{eq:asymrel} \Pr [i\in {\cal R}({\cal T}_{1},{\cal
T}_{2})]\sim q_{i}\expect [{\cal T}_{2}-{\cal T}_{1}]
\;\;\text{as}\;\;\text{$i\rightarrow \infty$,}
\end{equation}
where ${\cal R}(u,t)\eqdef {\cal R}_{1}(u,t)\cup \dots \cup{\cal
R}_{M}(u,t)$.\end{lemma} \proof Given in Section~\ref{sec:proofs}.
\qed

\section{Caching policies and the optimality}
\label{sec:main}

Consider infinitely many documents of unit size out of which $x$
can be stored in a local memory referred to as cache. When an item
is requested, the cache is searched first and we say that there is
a cache hit if the item is found in the cache. In this case the
cache content is left unchanged. Otherwise, we say that there is a
cache fault/miss and the missing item is brought in from the
outside world. At the time of a fault, a decision whether to
replace some item from the cache with a missing item has to be
made. We assume that replacements are optional, i.e., the cache
content can be left unchanged even in the case of fault. A caching
algorithm represents a set of document replacement rules.

We consider a class of caching algorithms whose information
decisions are made using only the information of past and present
requests and past decisions. More formally, let ${\cal
C}^{\pi}_{t}\equiv {\cal C}^{\pi}_t (x)$ be a cache content at
time $t$ under policy $\pi$. When the request for a document
$R_{n}$ is made, the cache with content ${\cal
C}^{\pi}_{\tau_{n}}$ is searched first. If document $R_{n}$ is
already in the cache ($R_{n}\in {\cal C}^{\pi}_{\tau_{n}}$), then
we use the convention that no document is replaced. On the other
hand, if document $R_{n}$ is not an element of ${\cal
C}^{\pi}_{\tau_{n}}$, then a document to be replaced is chosen
from a set ${\cal C}^{\pi}_{\tau_{n}}\cup\{R_{n}\}$ using a
particular eviction policy. At any moment of request, $\tau_n$,
the decision what to replace in the cache is based on $R_{1},
R_{2}, \dots, R_{n}, C^{\pi}_{\tau_{0}}, C^{\pi}_{\tau_{1}},\dots
,C^{\pi}_{\tau_{n}}$. Note that this information already contains
all the replacement decisions made up to time $\tau_{n}$. This is
the same information as the one used in the Markov decision
framework \cite{OBAM03}.

The set of the previously described cache replacement policies,
say ${\cal P}_{c}$, is quite large and contains mandatory caching
rules (more typical for a computer memory environment), i.e.,
those rules that require replacements in the case of cache faults.
Furthermore, the set ${\cal P}_{c}$ also contains the static
algorithm that places a fixed collection of documents $ {\cal
C}^{\pi}_{t}\equiv {\cal C}$ in the cache and then keeps the
same content without ever changing it.

Now, define the long-run cache fault probability corresponding to
the policy $\pi\in {\cal P}_{c}$ and a cache of size $x$ as
\begin{equation}
\label{eq:deffault1} P(\pi ,x)\eqdef
\limsup_{T\rightarrow\infty}\frac{\expect \left [\sum_{\tau_{n}\in
[0,T]} 1[R_n\not \in {\cal C}^{\pi}_{\tau_{n}}]\right ]}{T},
\end{equation}
recall that $\expect N(0,T)=T$. Note that we use the $\limsup$ in
this definition since the limit may not exist in general and that,
as defined before, ${\cal C}^{\pi}_{\tau_{n}} \equiv {\cal
C}^{\pi}_{\tau_n} (x)$ is a function of $x$ and we suppress it from the notation.

Next, we show that
\begin{equation}
\label{eq:deffault} P(\pi,x)=\limsup_{k\rightarrow
\infty}\frac{\expect \left [\sum_{\tau_{n}\in [0,{\cal
T}_{k}]}1[R_{n}\not \in {\cal C}^{\pi}_{\tau_{n}}]\right
]}{\expect N(0,{\cal T}_{k})},
\end{equation}
where ${\cal T}_k$ are the regenerative points, as defined in the
previous section. Note that estimating the previous expression is
not straight forward since replacement decision depends on all
previous requests, i.e., it depends on the past beyond the last
regenerative point. To this end, for the lower bound, for any
$0<\epsilon <1$, let $k\equiv k(T,\epsilon)\eqdef \lfloor
T(1-\epsilon)/\expect [{\cal T}_{2}-{\cal T}_{1}]\rfloor$, where
$\lfloor u \rfloor$ is the largest integer that is less or equal
to $u$. Then, note that
\begin{align}
\frac{1}{T}\expect \left [\sum_{\tau_{n}\in [0,T]} 1[R_{n}\not \in
{\cal C}^{\pi}_{\tau_{n}}]\right ]&\ge \expect \left [1[{\cal
T}_{k}<T]\frac{\sum_{\tau_{n}\in [0,{\cal T}_{k}]}1[R_{n}\not \in
{\cal C}^{\pi}_{\tau_{n}}]}{T}\right ]\nonumber\\
&\ge \expect \left [\frac{\sum_{\tau_{n}\in [0,{\cal
T}_{k}]}1[R_{n}\not \in {\cal C}^{\pi}_{\tau_{n}}]}{T}\right
]-\expect \left [1[{\cal T}_{k}>T]\frac{N(0,T)}{T}\right
]\label{eq:l100}.
\end{align}
Next, using the Weak Law of Large Numbers for $\Pr [{\cal
T}_{k}>T]\rightarrow 0$ (as $T\rightarrow \infty$) and the fact
that $N(0,T)$ is Poisson with mean $T$ in the preceding
inequality, we obtain
\begin{equation*}
P(\pi,x)\ge (1-\epsilon)\limsup_{\substack{T\rightarrow \infty\\
k=\left \lfloor \frac{T(1-\epsilon)}{\expect [{\cal T}_{2}-{\cal
T}_{1}]}\right \rfloor }} \frac{\expect \left [\sum_{\tau_{n}\in
[0,{\cal T}_{k}]}1[R_{n}\not \in {\cal C}^{\pi}_{\tau_{n}}]\right
]}{\expect N(0,{\cal T}_{k})}=(1-\epsilon)\limsup_{k\rightarrow
\infty} \frac{\expect \left [\sum_{\tau_{n}\in [0,{\cal
T}_{k}]}1[R_{n}\not \in {\cal C}^{\pi}_{\tau_{n}}]\right
]}{\expect N(0,{\cal T}_{k})},
\end{equation*}
since the set $\{k : k=\lfloor {T(1-\epsilon)}/{\expect [{\cal
T}_{2}-{\cal T}_{1}]}\rfloor, T>0\}$ covers all integers. We
complete the proof of the lower bound by passing
$\epsilon\rightarrow 0$. The upper bound uses similar arguments
where, in this case, $k$ is defined as $k\equiv
k(T,\epsilon)\eqdef \lfloor T(1+\epsilon)/\expect [{\cal
T}_{2}-{\cal T}_{1}]\rfloor$, and $P(\pi, x)$ is upper bounded as
\begin{equation*}
\frac{1}{T}\expect \left [\sum_{\tau_{n}\in [0,T]} 1[R_{n}\not \in
{\cal C}^{\pi}_{\tau_{n}}]\right ]\le \expect \left [1[T < {\cal
T}_{k}]\frac{\sum_{\tau_{n}\in [0,{\cal T}_{k}]}1[R_{n}\not \in
{\cal C}^{\pi}_{\tau_{n}}]}{T}\right ] + \expect \left [1[T >
{\cal T}_{k}]\frac{N(0,T)}{T}\right ].
\end{equation*}
Then, similarly to earlier arguments, we derive the
corresponding upper bound for $P(\pi, x)$ in (\ref{eq:deffault}).

Next, observe the static policy $s$, where ${\cal
C}^{\pi}_{\tau_{n}}\equiv \{1,2,\dots ,x\}$ for every $n$. Then,
due to the ergodicity of the request process, the long-run cache
fault probability of the static policy is
\begin{equation*}
P_{s}(x)\eqdef P(s,x)=\sum_{i>x}q_{i}.
\end{equation*}
Since the static policy belongs to
the set of caching algorithms ${\cal P}_{c}$, we conclude that
\begin{equation}
\label{eq:upbd4} P_{s}(x)\ge \inf_{\pi \in {\cal P}_{c}}P(\pi ,x).
\end{equation}

Our goal in this paper is to show that for large cache sizes $x$
there is no caching policy that performs better, i.e., achieves
long-term fault probability smaller than $P_{s}(x)$. This is
stated in the following main result of this paper.
\begin{theorem}
\label{theorem:main} For the semi-Markov modulated request process
defined in Section \ref{sec:model}, the static policy that stores
documents with the largest marginal popularities minimizes the
long-term cache fault probability for large caches, i.e.,
\begin{equation}
\label{eq:main} \inf_{\pi\in {\cal P}_{c}} P(\pi ,x)\sim
P_{s}(x)\;\;\text{as}\;\;\text{$x\rightarrow \infty$.}
\end{equation}
\end{theorem}

\noindent{\bf Remarks:} (i) From the examination of the following
proof it is clear that the result holds for any regenerative
request process that satisfies Lemma~\ref{lemma:asympr}. (ii)
Though asymptotically long-term optimal, the static frequency rule
possesses other undesirable properties such as high complexity and lack of 
adaptability to variations in the request patterns. However,
its optimal performance presents an important benchmark for
evaluating and comparing widely implemented caching policies in
the Web environment. On the other hand, it is a question whether a
widely accepted analysis of the cache miss ratio is the most
relevant performance measure to analyze. A strong argument in
support to this choice is that other measures would be harder
(sometimes impossible) to analyze. However, in Section
\ref{sec:concl}, we present some possible extensions of our
results to the analysis of other objective functions, such as
long-run average delay of fetching documents not found in the
cache, or long-run average cost of retrieving documents outside of
the cache, etc. (iii) Note that the condition $q_i$, $i\ge 1$,
given in the previous section makes the problem of proving
asymptotic optimality nontrivial. In case $q_i
> 0$ for just a finite number of $i$'s, the document population
would be finite and the result above would be trivially true. (iv)
The preliminary version of this work was presented in the Workshop
on Analytic Algorithms and Combinatorics (ANALCO'2006), Miami, Florida, January
2006.

\proof In view of (\ref{eq:upbd4}), we only need to show that $\inf_{\pi\in {\cal P}_{c}} P(\pi ,x)\gtrsim P_{s}(x)$
as $x\rightarrow \infty$.

For any set ${\mathcal A}$, let $|{\mathcal A}|$ denote the number
of elements in ${\mathcal A}$ and ${\mathcal A}\setminus {\mathcal
B}$ represent the set difference. Then, it is easy to see that the
number of cache faults in $[t,u)$, $t<u$, is lower bounded by
$|{\cal R}(t,u)\setminus {\cal C}_{t}^{\pi}|$ since every item
that was not in the cache at time $t$ results in at least one
fault when requested for the first time; in particular, if
$t={\cal T}_{j}$, $u={\cal T}_{j+1}$,
\begin{equation}
\label{eq:l5} \sum_{\tau_{n}\in [{\cal T}_{j},{\cal
T}_{j+1})}1[R_{n}\not \in {\cal C}^{\pi}_{\tau_{n}}]\ge |{\cal
R}({\cal T}_{j},{\cal T}_{j+1})\setminus {\cal C}^{\pi}_{{\cal
T}_{j}}|.
\end{equation}
This inequality and (\ref{eq:deffault}) results in
\begin{equation}
\label{eq:l200} P(\pi,x)\ge \limsup_{k\rightarrow
\infty}\frac{1}{\expect N(0,{\cal T}_{k})}\sum_{j=1}^{k-1}\expect
[ |{\cal R}({\cal T}_{j},{\cal T}_{j+1})\setminus {\cal
C}^{\pi}_{{\cal T}_{j}}|].
\end{equation}
Now, since we consider caching policies where replacement
decisions depend only on the previous cache contents and requests,
due to the renewal structure of the request process we conclude
that for every $j\ge 1$ and all $i\ge 1$, events $\{i\in {\cal
R}({\cal T}_{j},{\cal T}_{j+1})\}$ and $\{i\in {\cal
C}^{\pi}_{{\cal T}_{j}}\}$ are independent and, therefore, for
every $j\ge 1$,
\begin{align*}
\expect \left [|{\cal R}({\cal T}_{j},{\cal T}_{j+1})\setminus
{\cal C}^{\pi}_{{\cal T}_{j}}| 1[ {\cal C}_{{\cal
T}_{j}}^{\pi}={\cal C}]\right ]&= \sum_{i\ge 1} \Pr [i\in {\cal
R}({\cal T}_{j},{\cal T}_{j+1}),i\not \in {\cal C}]\Pr [{\cal
C}^{\pi}_{{\cal T}_{j}}={\cal
C}]\\
&=\Pr [{\cal C}^{\pi}_{{\cal T}_{j}}={\cal C}]\sum_{i\ge 1} \Pr
[i\in {\cal R}({\cal T}_{j},{\cal
T}_{j+1})]1[i\not\in {\cal C}]\\
&\ge \Pr [{\cal C}^{\pi}_{{\cal T}_{j}}={\cal C}]\inf_{{\cal
C}:|{\cal C}|=x}\sum_{i\not \in {\cal C}}\Pr [i\in {\cal R}({\cal
T}_{j},{\cal T}_{j+1})].
\end{align*}
Then, after summing over all values of ${\cal C}$, for any $j\ge
1$ we obtain
\begin{equation}
\label{eq:lowbd3} \expect [|{\cal R}({\cal T}_{j},{\cal
T}_{j+1})\setminus {\cal C}^{\pi}_{{\cal T}_{j}}|]\ge \inf_{{\cal
C}:|{\cal C}|=x} \sum_{i\not \in {\cal C}}\Pr [i\in {\cal R}({\cal
T}_{j},{\cal T}_{j+1})].
\end{equation}

Next, we show that the cache content ${\cal C}=[1,x]\eqdef \{1,\dots ,x\}$
achieves the infimum in the previous expression for large cache
sizes. This is equivalent to proving that, as $x\rightarrow
\infty$,
\begin{equation}
\label{eq:lowbdc2}
 \inf_{{\cal C}:|{\cal C}|=x} \sum_{i\not \in {\cal
C}}\Pr [i\in {\cal R}({\cal T}_{j},{\cal T}_{j+1})]\gtrsim
\sum_{i\not \in [1,x]}\Pr [i\in {\cal R}({\cal T}_{j},{\cal
T}_{j+1})].
\end{equation}
We will justify the previous statement by showing that for any set
${\cal C}$ obtained from $[1,x]$ by placing documents from the set
$\{x+1,\dots\}$ instead of those in $[1,x]$ can not result in
$\sum_{i\not \in {\cal C}}\Pr [i\in {\cal R}({\cal T}_{j},{\cal
T}_{j+1})]< (1-\epsilon)\sum_{i\not \in [1,x]} \Pr [i\in {\cal
R}({\cal T}_{j},{\cal T}_{j+1})]$ for large cache sizes $x$ and
any $0<\epsilon <1$.

Lemma~\ref{lemma:asympr} implies that for an
arbitrarily chosen $\epsilon>0$ there exists finite integer
$i_{0}$ such that for all $i\ge i_{0}$
\begin{equation}
(1-\epsilon)q_{i}\expect [{\cal T}_{j+1}-{\cal T}_{j}]< \Pr [i\in
{\cal R}({\cal T}_{j},{\cal T}_{j+1})]< (1+\epsilon)q_{i}\expect
[{\cal T}_{j+1}-{\cal T}_{j}].\label{eq:bounds}
\end{equation}
Thus, using the previous expression and $q_{i}\downarrow 0$ as
$i\rightarrow \infty$, we conclude that for all $k\le i_{0}$ there
exists $x_{0}\ge i_{0}$, such that for all   $i\ge x_{0}$
\begin{equation}
\label{eq:lessxst} \min_{1\le k\le i_{0}}\Pr [k\in {\cal R}({\cal
T}_{j},{\cal T}_{j+1})]> \Pr [i\in {\cal R}({\cal T}_{j},{\cal
T}_{j+1})].
\end{equation}
Now, assume that the cache is of size $x\ge x_{0}$ and observe
different cache contents ${\cal C}$ obtained from $[1,x]$ by
replacing its documents with items from $\{x+1,x+2,\dots\}$. Next,
using (\ref{eq:lessxst}), we conclude that replacing documents
enumerated with $\{1,\dots ,i_{0}\}$ can only increase the sum on
the left-hand side of (\ref{eq:lowbdc2}). On the other hand,
observe cache contents ${\cal C}$ that are obtained from $[1,x]$
by replacing documents enumerated as $\{i_{0}+1,\dots ,x\}$ with
items from $\{x+1,\dots\}$. Then, it is easy to see that proving
inequality (\ref{eq:lowbdc2}) is equivalent to showing that
$\sum_{i\in[i_{0}+1,x]}\Pr [i\in {\cal R}({\cal T}_{j},{\cal
T}_{j+1})]\ge (1-\epsilon)\sum_{i\in {\cal C}\setminus
[1,i_{0}]}\Pr [i\in {\cal R}({\cal T}_{j},{\cal T}_{j+1})]$, for
any $0<\epsilon<1$. Next, since for any $i\ge i_{0}$ inequalities
(\ref{eq:bounds}) hold, we conclude
\begin{equation*}
\frac{\sum_{i\in [i_{0}+1,x]}\Pr [i\in {\cal R}({\cal T}_{j},{\cal
T}_{j+1})]}{\sum_{i\in {\cal C}\setminus [1,i_{0}]}\Pr [i\in{\cal
R}({\cal T}_{j},{\cal
T}_{j+1})]}\ge\frac{(1-\epsilon)\sum_{i\in[i_{0}+1,x]}q_{i}}{(1+\epsilon)\sum_{i\in
{\cal C}\setminus [1,i_{0}]}q_{i}}\ge
\frac{1-\epsilon}{1+\epsilon},
\end{equation*}
where the second inequality in the previous expression follows
from the monotonicity of $q_{i}$s. Then, by passing $\epsilon\rightarrow 0$ we prove
inequality (\ref{eq:lowbdc2}).

Note that after applying the lower bound (\ref{eq:lowbdc2}) in
(\ref{eq:lowbd3}), in conjunction with (\ref{eq:l200}), the
renewal nature of the regenerative points and
Lemma~\ref{lemma:asympr}, we obtain that, as $x\rightarrow
\infty$,
\begin{equation}
\label{eq:finexp} \inf_{\pi\in {\cal P}_c} P(\pi,x)\gtrsim
\sum_{i\ge x}q_{i},
\end{equation}
which completes the proof of the theorem.
\qed

\section{Further extensions and concluding remarks}
\label{sec:concl}

In this paper we prove that the static frequency rule minimizes
the long term fault probability in the presence of correlated requests for large cache sizes.

There are several generalizations of our results that are worth
mentioning. First, the definition of the fault probability in
(\ref{eq:deffault}) can be generalized by replacing terms
$1[R_{n}\not \in {\cal C}^{\pi}_{\tau_{n}}]$ with
$f(R_{n})1[R_{n}\not \in {\cal C}^{\pi}_{\tau_{n}}]$, where $f(i)$
could represent the cost of retrieving document $i$, e.g., the
delay of fetching item $i$ not found in the cache. Assume that
$0<f(i)\le K<\infty$ and let ${\cal S}$ be a set of $x$ items such that
$q_i f(i)\ge q_j f(j)$ for all $i\in {\cal S}$ and $j\not\in {\cal S}$.  Then, the following result holds:
\begin{theorem}
\label{eq:delay} For the semi-Markov modulated request process
defined in Section~\ref{sec:model}, the static caching policy
${\cal C}\equiv {\cal S}$ minimizes the long-run average cost function $f(\cdot)$ (e.g., delay)
for documents not found in the cache.
\end{theorem}
\noindent{\bf Sketch of the proof:} The proof of this theorem
follows completely analogous arguments to those used in the proof
of Theorem \ref{theorem:main}, and, in order to avoid repetitions,
we outline its basic steps.

Similarly as in (\ref{eq:deffault1}), the long-run average cost
for documents not found in the cache that corresponds to
the caching policy $\pi\in {\cal P}_c$ is defined as
\begin{equation*}
D(\pi,x)\eqdef \limsup_{T\rightarrow\infty}\frac{\expect \left
[\sum_{\tau_{n}\in [0,T]} f(R_n)1[R_n\not \in {\cal
C}^{\pi}_{\tau_{n}}]\right ]}{T}.
\end{equation*}
Then, by using similar arguments to (\ref{eq:deffault}) -
(\ref{eq:upbd4}) and $0<f(i)\le K<\infty$, $i\ge 1$, we obtain
that the long-run average cost of the static policy ${\cal
C}_{\tau_n}\equiv {\cal S}$, $n\ge 1$, for the cache with size $x$
satisfies
\begin{equation}
\label{eq:dod1} D_s(x)=\sum_{i\not \in {\cal S}} f(i)q_i\ge
\inf_{\pi\in {\cal P}_c}D(\pi,x).
\end{equation}

Next, in order to prove
\begin{equation}
\label{eq:dod2} D_s(x)\lesssim \inf_{\pi\in {\cal
P}_c}D(\pi,x)\;\;\text{as $x\rightarrow \infty$,}
\end{equation}
similarly as in the proof of Theorem \ref{theorem:main}, we lower
bound the number of cache misses, and, therefore, the average cost in
every regenerative interval $[{\cal T}_j,{\cal T}_{j+1})$, $j\ge
1$, as
\begin{equation*}
\sum_{\tau_n \in [{\cal T}_j,{\cal T}_{j+1})} f(R_n)1[R_n\not \in
{\cal C}^{\pi}_{\tau_n}]\ge \sum_{i\ge 1}f(i)1[i\in {\cal R}({\cal
T}_j,{\cal T}_{j+1}),i\not \in {\cal C}_{{\cal T}_j}].
\end{equation*}
Next, since we consider caching policies whose replacement
decisions depend only on the past cache contents and requests, due
to the renewal structure of the request process, we conclude that
for any $j\ge 1$,
\begin{equation*}
\expect \left [\sum_{\tau_n\in [{\cal T}_j,{\cal T}_{j+1})}
f(R_n)1[R_n\not \in {\cal C}^{\pi}_{\tau_n}]1[{\cal
C}^{\pi}_{{\cal T}_j}={\cal C}]\right ]\ge \Pr [{\cal
C}^{\pi}_{{\cal T}_j}={\cal C}]\sum_{i\not \in {\cal C}} f(i)\Pr
[i\in {\cal R}({\cal T}_j,{\cal T}_{j+1})]
\end{equation*}
and, thus, similarly as in (\ref{eq:lowbd3}), we obtain
\begin{equation*}
\expect \left [\sum_{\tau_n\in [{\cal T}_j,{\cal T}_{j+1})}
f(R_n)1[R_n\not \in {\cal C}^{\pi}_{\tau_n}]\right ]\ge
\inf_{{\cal C}:|{\cal C}|=x}\sum_{i\not \in {\cal C}}f(i)\Pr [i\in
{\cal R}({\cal T}_j,{\cal T}_{j+1})].
\end{equation*}
Now, given the previous observations, the asymptotic inequality
(\ref{eq:dod2}) is proved using analogous arguments to those in
(\ref{eq:bounds}) - (\ref{eq:finexp}). Note that in the context of
this result, inequality (\ref{eq:lessxst}) becomes
\begin{equation*}
\min_{1\le k\le i_0} f(k)\Pr [k\in {\cal R}({\cal T}_j,{\cal
T}_{j+1})]>f(i)\Pr [i\in {\cal R}({\cal T}_j,{\cal T}_{j+1})]
\end{equation*}
for all $i\ge x_0$, and we have analogous asymptotic linearity as
in (\ref{eq:bounds}) since $q_i\downarrow 0$ as $i\rightarrow
\infty$ and $0<f(i)\le K<\infty$. Finally, as in
(\ref{eq:lessxst}) - (\ref{eq:finexp}), the rest of the proof is
based on proving that no replacements of documents in the set
${\cal S}$ can lead to smaller long-run average delays for large
cache sizes $x$. Thus, the asymptotic bound (\ref{eq:dod2}) holds
and, in conjunction with (\ref{eq:dod1}), completes the proof of
the theorem. \qed

In addition to the previous generalization, in the context of
documents with different sizes, one can prove the following
result:
\begin{theorem}
Assume that documents have different sizes and that they are
enumerated according to the non-increasing order of $q_i/s_i$,
i.e., $q_1/s_1\ge q_2/s_2\ge \dots$, where $s_i$ is the size of
document $i$ and $s_i\in \{s_1,\dots ,s_D\}$, where $s_1,\dots
,s_D<\infty$ and $D<\infty$. Then, for the semi-Markov modulated
request process defined in Section \ref{sec:model}, if $\sum_{j>i}
q_j\sim \sum_{j\ge i}q_j$ as $i\rightarrow \infty$, i.e.,
$\sum_{j>i} q_j$ is long-tailed, the static rule that places
documents with the smallest index in the cache, subject to the
constraint $\sum_i s_i\leq x$, is asymptotically optimal.
\end{theorem}
\proof In light of the identical arguments to those used in the
proof of Theorem \ref{theorem:main}, it is not hard to show that
the static policy minimizes the long run average number of misses
for large cache sizes. More specifically, the optimal long-run
cache fault probability is of the form
\begin{equation}
\label{eq:probloss} P_s (x)=\sum_{i\in {\cal N}_x} q_i,
\end{equation}
where ${\cal N}_x$ is the set of document indices that minimizes
(\ref{eq:probloss}) subject to the constraint $\sum_{i\not \in
{\cal N}_x}s_i\le x$. The previous problem is a knapsack problem
(see Section 5.2 of \cite{MOS91} for further explanations). It is
shown that in the case where objects can be split to exactly fill
the knapsack, the policy that minimizes (\ref{eq:probloss}) is the
one that places documents with the largest $q_i/s_i$ values in the
cache until an object, say $n_x$, fails to fit. Then, the
optimal solution is to split document $n_x$ to fill the
cache completely. Since that is not possible in the case of
document caching, it is not hard to see that the fault probability
for the optimal static placement in our case is between the
optimal fault probabilities in the case of cache sizes
$\sum_{i=1}^{n_{x}-1}s_i$ and $\sum_{i=1}^{n_x}s_i$ (note that
$\sum_{i=1}^{n_{x}-1}s_i\le x < \sum_{i=1}^{n_x}s_i$). Now, since
$n_x$ monotonically increases as $x$ increases, in conjunction
with the long-tailed assumption of the theorem, $\sum_{i\ge n_x} q_i\sim
\sum_{i>n_x} q_i$ as $x\rightarrow \infty$, we conclude the proof
of the theorem. \qed

Finally, in light of our recent result on the asymptotic
performance of the ordinary LRU caching rule in the presence of
semi-Markov modulated requests and Zipf's law marginal
distributions ($q_{i}\sim c/i^{\alpha}$ as $i\rightarrow \infty$,
$c>0$) obtained in Theorem~3 of \cite{JERA02}, asymptotic
optimality of the static frequency rule implies that the LRU is
factor $e^{\gamma}\approx 1.78$ away from the optimal ($\gamma$ is
the Euler constant, i.e. $\gamma \approx 0.57721\dots$).
Therefore, in view of other desirable properties, its
self-organizing nature and low complexity, the LRU rule has
excellent performance even in the presence of statistically
correlated requests. Furthermore, stronger optimality
conclusions could be drawn for the recently proposed versions of
the LRU policy (see \cite{JERAX05} and \cite{JERA08}), given that
the performance analysis of these algorithms can be extended to
the correlated setting such as the one in \cite{JERA02}.

\section{Proof of Lemma
\ref{lemma:asympr}}
\label{sec:proofs}

In this section, we prove the asymptotic relation
(\ref{eq:asymrel}) stated at the end of Section \ref{sec:model}.

Note that
\begin{align}
\Pr [i\in {\cal R}({\cal T}_{1},{\cal T}_{2})]&=1-\Pr [i\not \in {\cal R}_{1}({\cal T}_{1},{\cal T}_{2}),\dots
,i\not \in {\cal R}_{M}({\cal T}_{1},{\cal
T}_{2})]\nonumber\\
&=\expect \left [1-(1-q_{i}^{(1)})^{N_{1}}\dots
(1-q_{i}^{(M)})^{N_{M}}\right ],\label{eq:eqexpr1}
\end{align}
where $N_{r}\eqdef N_{r}({\cal T}_{1},{\cal T}_{2})$, $1\le r\le
M$. Then, since $q_{i}\rightarrow 0$ as $i\rightarrow \infty$ and $\min_r \pi_r > 0$, it
follows that $q_{i}^{(r)}\rightarrow 0$ as $i\rightarrow \infty$,
$1\le r\le M$. In addition, $1-e^{-x}\le x$ for all $x\ge 0$ and
for any $1>\epsilon >0$, there exists $x_{0}(\epsilon)>0$, such
that for all $0\le x\le x_{0}(\epsilon)$ inequality $1-x\ge
e^{-x(1+\epsilon)}$ holds, and, therefore, for $i$ large enough
\begin{align}
\expect \left [1-e^{-(q_{i}^{(1)}N_{1}+\dots
+q_{i}^{(M)}N_{M})}\right ]&\le \expect \left
[1-(1-q_{i}^{(1)})^{N_{1}}\dots (1-q_{i}^{(M)})^{N_{M}}\right
]\nonumber\\
&\le \expect \left [1-e^{-(1+\epsilon)(q_{i}^{(1)}N_{1}+\dots
+q_{i}^{(M)}N_{M})}\right ].\label{eq:secline}
\end{align}
Then, since $1-e^{-x}\le x$ for
$x\ge 0$, we obtain, for $i$ large
enough,
\begin{equation}
\expect\left
[1-e^{-(1+\epsilon)(N_{1}q_{i}^{(1)}+\dots+N_{M}q_{i}^{(M)})}\right
]\le(1+\epsilon) \expect \left
[q_{i}^{(1)}N_{1}+\dots+q_{i}^{(M)}N_{M}\right
].\label{eq:firstterm}
\end{equation}

Next, let $N\eqdef N_{1}+\dots +N_{M}$. Then, we show that
$q_{i}^{(1)}\expect N_{1}+\dots+q_{i}^{(M)}\expect
N_{M}=q_{i}\expect N$. From the ergodicity of $J_{t}$,
it follows that
\begin{equation*}
\Pr [J_t = r]=\frac{\expect {\cal T}_{1r}}{\expect [{\cal
T}_{2}-{\cal T}_{1}]},
\end{equation*}
where ${\cal T}_{1r}$, $1\le r\le M$, is the length of time that
$J_{t}$ spends in state $r$ during the renewal interval $({\cal
T}_{1},{\cal T}_{2})$ (see Section~1.6 of \cite{FBPB02}). Finally,
using $\expect N=\expect[{\cal T}_{2}-{\cal T}_{1}]$ and $\expect
N_{r}=\expect {\cal T}_{1r}$, $1\le r\le M$ (Poisson process of
rate $1$), in conjunction with (\ref{eq:firstterm}), we conclude,
for $i$ large
\begin{equation}
\label{eq:firstasym} \expect\left
[1-e^{-(1+\epsilon)(N_{1}q_{i}^{(1)}+\dots
+N_{M}q_{i}^{(M)})}\right ]\le(1+\epsilon)q_{i}\expect [{\cal
T}_{2}-{\cal T}_{1}].
\end{equation}

Next, we estimate the lower bound for the left hand side in
(\ref{eq:secline}). After conditioning, we obtain
\begin{equation}
\expect \left [1-e^{-(N_{1}q_{i}^{(1)}+\dots
+N_{M}q_{i}^{(M)})}\right ]\ge \expect \left [\left (1-e^{-(N_{1}q_{i}^{(1)}+\dots
+N_{M}q_{i}^{(M)})}\right )1\left [N\le \Bar
{q}_{i}^{-\frac{1}{2}}\right ]\right ],\label{eq:thirdline}
\end{equation}
where $q_{i}^{(r)}\le \Bar {q}_{i}\eqdef q_{i}/\min_{r}\pi_{r}\le
Hq_{i}$, $1\le r\le M$, and some large enough constant
$0<H<\infty$ . Then, note that for every $\omega \in \{ N\le \Bar
{q}_{i}^{-\frac{1}{2}}\}$, $q_{i}^{(1)}N_{1}+\dots +
q_{i}^{(M)}N_{M}\le \frac{\Bar {q}_{i}}{\sqrt{\Bar {q}_{i}}}=\sqrt
{\Bar {q}_{i}}$. In addition, for any $1>\epsilon
>0$, there exists $x_{\epsilon}>0$, such that for all $0\le
x\le x_{\epsilon}$ inequality $1-e^{-x}\ge (1-\epsilon)x$ holds
and, therefore, for $i$ large enough such that $\sqrt {\Bar
{q}_{i}}\le x_{\epsilon}$
\begin{align*}
\expect \left [ \left (1-e^{-(N_{1}q_{i}^{(1)}+\dots
+N_{M}q_{i}^{(M)})}\right )1\left [N\le \Bar
{q}_{i}^{-\frac{1}{2}}\right ]\right ]&\ge (1-\epsilon)\expect
\left [(N_{1}q_{i}^{(1)}+\dots +N_{M}q_{i}^{(M)})1\left [N\le \Bar
{q}_{i}^{-\frac{1}{2}}\right ]\right ]\\
&\ge (1-\epsilon)q_{i}\expect [{\cal T}_{2}-{\cal
T}_{1}]-(1-\epsilon)\Bar {q}_{i}\expect \left [N1 \left [N>\Bar
{q}_{i}^{-\frac{1}{2}}\right ]\right ].
\end{align*}
Then, since $\expect N<\infty$ and $1/\sqrt {\Bar
{q}_i}\rightarrow \infty$ as $i\rightarrow \infty$, it is
straightforward to conclude that
 $\expect [N1[N>1/\sqrt{\Bar {q}_{i}}]]= \expect N - \expect[N1[N\le 1/\sqrt{\Bar {q}_{i}}]]\rightarrow 0$ as
$i\rightarrow \infty$, and, therefore, in conjunction with
(\ref{eq:thirdline}), we obtain
\begin{equation*}
\expect \left [1-e^{-(N_{1}q_{i}^{(1)}+\dots
+N_{M}q_{i}^{(M)})}\right ]\gtrsim (1-\epsilon)q_{i}\expect [{\cal
T}_{2}-{\cal T}_{1}],
\end{equation*}
as $i\rightarrow \infty$. Finally, after letting
$\epsilon\rightarrow 0$ in the previous expression and
(\ref{eq:firstasym}), we complete the proof of this lemma. \qed

\section*{Acknowledgements}
We thank an anonymous reviewer for his/her helpful comments.


\begin{thebibliography}{10}
\bibitem{WA97}
M.~Abrams and R.~Wooster.
\newblock Proxy caching that estimates edge load delays.
\newblock In {\em Proceedings of 6th {I}nternational {W}orld {W}ide {W}eb {C}onference},
  Santa Clara, CA, April 1997.

\bibitem{ABC96}
V.~Almeida, A.~Bestavros, M.~Crovella, and A.~de~Oliviera.
\newblock Characterizing reference locality in the {WWW}.
\newblock In {\em Proceedings of the Fourth International Conference on
  Parallel and Distributed Information Systems}, Miami Beach, Florida, December
  1996.

\bibitem{AW96}
M.~Arlitt and C.~Williamson.
\newblock Web server workload characteristics: {T}he search for invariants.
\newblock In {\em Proceedings of {ACM} {SIGMETRICS}'1996}, Philadelphia, PA, May
  1996.

\bibitem{FBPB02}
F.~Baccelli and P.~Br\'emaud.
\newblock {\em Elements of Queueing Theory}.
\newblock Springer--Verlag, 2002.

\bibitem{OBAM03}
O.~Bahat and A.~M. Makowski.
\newblock Optimal replacement policies for non-uniform cache objects with
  optional eviction.
\newblock In {\em Proceedings of {IEEE INFOCOM}'2003}, San Francisco,
  California, USA, April 2003.

\bibitem{BCF99}
L.~Breslau, P.~Cao, L.~Fan, G.~Phillips, and S.~Shenker.
\newblock Web caching and {Z}ipf-like distributions: Evidence and implications.
\newblock In {\em Proceedings of {IEEE INFOCOM}'1999}, New York, NY, March
  1999.

\bibitem{CAI97}
P.~Cao and S.~Irani.
\newblock Cost-aware {WWW} proxy caching algorithms.
\newblock In {\em Proceedings of the {USENIX}'1997 Annual Technical
  Conference}, Anaheim, California, January 1997.

\bibitem{BDxx}
Brian~D. Davison.
\newblock Web {C}aching and {C}ontent {D}elivery {R}esources.
\newblock In {\em http://www.web-caching.com}.

\bibitem{JERA02a}
P.~R. Jelenkovi\'c and A.~Radovanovi\'c.
\newblock Asymptotic {I}nsensitivity of {L}east-{R}ecently-{U}sed {C}aching to
  {S}tatistical {D}ependency.
\newblock In {\em Proceedings of {IEEE INFOCOM}'2003}, San Francisco, April
  2003.

\bibitem{JERA02}
P.~R. Jelenkovi\'c and A.~Radovanovi\'c.
\newblock Least-{R}ecently-{U}sed {C}aching with {D}ependent {R}equests.
\newblock {\em Theoretical Computer Science}, 326(1-3):293--327, 2004.

\bibitem{JERAX05}
P.~R. Jelenkovi\'c, X.~Kang, and A.~Radovanovi\'c.
\newblock Near optimality of the discrete persistent access caching algorithm.
\newblock {\em Discrete Mathematics and Theoretical Computer Science},
  AD:201--222, 2005.

\bibitem{JERA08}
P.~R. Jelenkovi\'c and A.~Radovanovi\'c.
\newblock The {P}ersistent-{A}ccess-{C}aching {A}lgorithm.
\newblock {\em Random Structures and Algorithms}, 33(2):219--251, May 2008.


\bibitem{JB00a}
S.~Jin and A.~Bestavros.
\newblock {G}reedy{D}ual* {W}eb {C}aching {A}lgorithm.
\newblock In {\em Proceedings of the 5th {I}nternational {W}eb {C}aching and
  {C}ontent {D}elivery {W}orkshop}, Lisbon, Portugal, May 2000.

\bibitem{JB00}
S.~Jin and A.~Bestavros.
\newblock Sources and characteristics of {W}eb temporal locality.
\newblock In {\em Proceedings of {M}ascots'2000: {T}he {IEEE/ACM}
  {I}nternational {S}ymposium on {M}odeling, {A}nalysis and {S}imulation of
  {C}omputer and {T}elecommunication {S}ystems}, San Fransisco, CA, August
  2000.

\bibitem{COD73}
E.~G.~Coffman Jr. and P.~J. Denning.
\newblock {\em Operating Systems Theory}.
\newblock Prentice-Hall, 1973.

\bibitem{MOS91}
B.~Moret and H.~Shapiro.
\newblock {\em Algorithms from P to NP: Volume 1 Design and Efficiency}.
\newblock The Benjamin/Cummings Publishing Company, Redwood City, CA, 1991.


\end{thebibliography}
\end{document}